\documentclass[twoside,twocolumn,9pt]{article}
\usepackage{extsizes}
\usepackage[super,sort&compress,comma]{natbib} 
\usepackage[version=3]{mhchem}
\usepackage[left=1.5cm, right=1.5cm, top=1.785cm, bottom=2.0cm]{geometry}
\usepackage{balance}
\usepackage{mathptmx}
\usepackage{sectsty}
\usepackage{graphicx} 
\usepackage{lastpage}
\usepackage[format=plain,justification=justified,singlelinecheck=false,font={stretch=1.125,small,sf},labelfont=bf,labelsep=space]{caption}
\usepackage{float}
\usepackage{fancyhdr}
\usepackage{fnpos}
\usepackage[english]{babel}
\addto{\captionsenglish}{%
  
}
\usepackage{array}
\usepackage{droidsans}
\usepackage{charter}
\usepackage[T1]{fontenc}
\usepackage[usenames,dvipsnames]{xcolor}
\usepackage{setspace}
\usepackage[compact]{titlesec}
\usepackage{hyperref}
\usepackage{ulem}

\usepackage{epstopdf}

\definecolor{cream}{RGB}{222,217,201}

\begin{document}

\pagestyle{fancy}
\thispagestyle{plain}
\fancypagestyle{plain}{
\renewcommand{\headrulewidth}{0pt}
}

\makeFNbottom
\makeatletter
\renewcommand\LARGE{\@setfontsize\LARGE{15pt}{17}}
\renewcommand\Large{\@setfontsize\Large{12pt}{14}}
\renewcommand\large{\@setfontsize\large{10pt}{12}}
\renewcommand\footnotesize{\@setfontsize\footnotesize{7pt}{10}}
\makeatother

\renewcommand{\thefootnote}{\fnsymbol{footnote}}
\renewcommand\footnoterule{\vspace*{1pt}%
\color{cream}\hrule width 3.5in height 0.4pt \color{black}\vspace*{5pt}} 
\setcounter{secnumdepth}{5}

\makeatletter 
\renewcommand\@biblabel[1]{#1}            
\renewcommand\@makefntext[1]%
{\noindent\makebox[0pt][r]{\@thefnmark\,}#1}
\makeatother 
\renewcommand{\figurename}{\small{Fig.}~}
\sectionfont{\sffamily\Large}
\subsectionfont{\normalsize}
\subsubsectionfont{\bf}
\setstretch{1.125} 
\setlength{\skip\footins}{0.8cm}
\setlength{\footnotesep}{0.25cm}
\setlength{\jot}{10pt}
\titlespacing*{\section}{0pt}{4pt}{4pt}
\titlespacing*{\subsection}{0pt}{15pt}{1pt}

\fancyfoot{}
\fancyfoot[RO]{\footnotesize{\sffamily{1--\pageref{LastPage} ~\textbar  \hspace{2pt}\thepage}}}
\fancyfoot[LE]{\footnotesize{\sffamily{\thepage~\textbar\hspace{3.45cm} 1--\pageref{LastPage}}}}
\fancyhead{}
\renewcommand{\headrulewidth}{0pt} 
\renewcommand{\footrulewidth}{0pt}
\setlength{\arrayrulewidth}{1pt}
\setlength{\columnsep}{6.5mm}
\setlength\bibsep{1pt}

\makeatletter 
\newlength{\figrulesep} 
\setlength{\figrulesep}{0.5\textfloatsep} 

\newcommand{\topfigrule}{\vspace*{-1pt}%
\noindent{\color{cream}\rule[-\figrulesep]{\columnwidth}{1.5pt}} }

\newcommand{\botfigrule}{\vspace*{-2pt}%
\noindent{\color{cream}\rule[\figrulesep]{\columnwidth}{1.5pt}} }

\newcommand{\dblfigrule}{\vspace*{-1pt}%
\noindent{\color{cream}\rule[-\figrulesep]{\textwidth}{1.5pt}} }

\makeatother

\twocolumn[
  \begin{@twocolumnfalse}
\vspace{1em}
\sffamily
\begin{tabular}{m{0.5cm} p{13.5cm} }

~ & \noindent\LARGE{\textbf{Comment on Boosting large scale capacitive harvesting of osmotic power by dynamical matching of ion exchange kinetics.}} \\

 & \noindent\large{Nan Wu \textit{$^{a}$}, Mathieu BA Freville,\textit{$^{a}$}, Zhiyi Man,\textit{$^{a}$}, Adérito Fins Carreira,\textit{$^{a}$}, Timothée Derkenne\textit{$^{a}$},  Corentin Tregouet,\textit{$^{a}$} and Annie Colin \textit{$^{\ast}$$^{a}$}}\\

~ & \noindent\normalsize{This article is a  comment of N. Chapuis and L. Bocquet, Sustainable Energy Fuels, 2025, DOI: 10.1039/D4SE01366B which title is "Boosting large scale capacitive harvesting of osmotic power by dynamical matching of ion exchange kinetics". In this work, the authors present an experimental process that shows how it is possible to set up a reverse electrodialysis cell capable of achieving power values of 5 W.m$^{-2}$. This value is the profitability threshold. Our work challenges this claim and questions whether the proposed technique can be scaled up, as well as the modeling of the process.} 

\end{tabular}

 \end{@twocolumnfalse} \vspace{0.6cm}

]

\renewcommand*\rmdefault{bch}\normalfont\upshape
\rmfamily
\section*{}
\vspace{-1cm}


\footnotetext{\textit{$^{a}$~ MIE – Chemistry, Biology and Innovation (CBI) UMR8231, ESPCI Paris, CNRS, PSL Research University, 10 rue Vauquelin, Paris, France}}
\footnotetext{\textit{$^{\ast}$~annie.colin@espci.psl.eu }}




\section{Blue Energy and the Challenges of Scalability}

Blue energy, also known as osmotic or salinity gradient energy, has been studied since Pattle's work in 1954\cite{pattle1954production}. Various technologies have been explored to harness this energy, including pressure retarded osmosis (PRO) and reverse electrodialysis (RED). However, the field lacks a clear scientific and technical consensus. 

Over the past decades, extensive research has focused on measurements at the scale of single nanopores or submillimeter membranes. These studies have reported power densities ranging from 20 to 100 W.m$^{-2}$, sparking growing interest in potential applications\cite{Guo2010SingleNanopore,Feng2016MOS2,Siria2013Giant}. However, they have not significantly advanced practical developments. Unlike many applied fields where an understanding of the nanometric or microscopic scale is fundamental, the emphasis here must be on the process scale, i.e. the centimeter or meter scale. 
This is because the results obtained at the nanometric scale cannot be reproduced in larger systems. Studies on single pores do not allow us to predict behavior at the membrane level, because pore-pore interactions considerably modify performance when the pore density is high. In fact, the performance of an N-pore membrane is not simply N times greater than that of a single pore, it is considerably lower\cite{gaoUnderstandingGiantGap2019}.

The study of model membranes with a surface area of a few hundred square microns poses significant measurement problems. They are often tested using centimeter-sized electrodes, which causes curvature of the field lines and focusing effects. Consequently, the observed response is not directly proportional to the surface area of the membrane and the properties measured are not relevant for applications, contrary to what is constantly reported in the literature\cite{derkenneMacroscopicAccessResistances2024}.
It is therefore crucial to employ microfluidic and fluidic technologies and work at larger scales to obtain results that are both realistic and applicable. This approach has been adopted in the study by Bocquet and Chapuis, which presents a capacitive technology with a reported performance that exceeds 5 W.m$^{-2}$\cite{chapuisBoostingLargeScale2025}. 

Capacitive technologies offer several advantages, particularly the absence of overpotential at the electrodes and the elimination of toxic chemicals, making them an attractive alternative to traditional electrochemical methods. The general principle of the capacitive concentration cell with a single membrane has already been described extensively by Brahmi et al\cite{brahmiNewMembraneElectrode2022}. In this setup, a positively charged ion-exchange membrane (I-CEM) is placed at the center of the cell, separating two compartments with different salinities. The membrane selectively allows cations to pass while blocking anions. This selectivity generates an electrical potential difference between the two solutions. When the circuit is closed, an electric current flows due to the potential drop, but the current decays over time as the capacitor charges. Once fully charged, the potential difference across its terminals balances the membrane potential $E_{mem}$ and the potential created by the adsorbed layers $E_{elec}$. At this point, fresh and saltwater supplies must be switched to reverse the membrane and electrode potential, allowing electricity production to resume in the opposite direction. Beyond salinity gradient energy recovery, capacitive devices are also being developed for CO$_2$ capture, expanding their potential applications. 

At first glance, the study appears to be highly significant due to the higher reported power density compared to previous studies (such as Brahmi et al.\cite{brahmiNewMembraneElectrode2022}, Zhu et al.\cite{zhuCarbonizedPeatMoss2019} and Zhan et al.\cite{zhanHighPerformanceConcentration2018}), and to the claim of scalability of the device. 

However, a closer analysis reveals potential oversights in essential considerations when evaluating the system's efficiency. The authors emphasize the scalability and commercial viability of their results, which makes it essential to realistically assess the practical potential of the system.

In the following, we detail our main sources of concern. In Section \ref{power}, we revisit the calculation of output power, a key claim of the article, and show that it may not be accurate. In particular, we find that the reported value should be negative, as the device appears to consume significantly more energy in viscous losses than it produces through the salt gradient. Then, in Section \ref{scalability}, we examine the claim of scalability, which is central to the article, and highlight some challenges that arise even when disregarding the previously discussed viscous losses.

\section{Validity of the reported power output and energy-conversion efficiency}
\label{power}
 The authors claim that their results significantly outperform those in the existing literature. The originality of their approach is based on two points. The first is assembling the membrane and conductive felt under high mechanical pressure to reduce contact resistance. This goal is perfectly achieved because the electrical resistance of the device is 10 times less than the average resistance of other devices.
The second original feature of the work concerns the use of very rapid cycles with short periods. This obviously makes it possible to recover power close to the maximum instantaneous power.
These two approaches require very high flow rates. The flow velocity must be increased to quickly impose the value of the salt concentration in the compartment on the membrane when a dense felt is compressed on the membrane.  A high flow velocity must also be imposed so that the period of change in concentration is greater than the filling time of the cell.

This approach therefore leads the authors to use flow rates that are 7 to 10 times greater than those in the literature. These high flow rates result in important viscous losses. Given the experimental setup and the system studied, these losses are likely to be significant and cannot be overlooked. 
In further details, it is well-established that the energy efficiency of any energy conversion process is defined by the following relation :
\begin{equation}
\mathcal{P}_{Net} = \mathcal{P}_{Harvested} - \mathcal{P}_{Dissipated} 
\end{equation}

The description of the flow geometry in the article seems not fully described. We assumed the following: the graphite current collector contains a narrow channel measuring 1.5 mm in width, 2.0 cm in length, and 1.5 mm in thickness. This channel is separated from the membrane by a larger channel (1.0 cm wide), which is also filled with electrode material. We investigated two scenarios regarding this second channel: A- one based on Figure 1b of the article, where there is a gasket of unknown thickness that we therefore assume (in an optimistic scenario) to be 1.0 mm thick; B- one based on the text, in which we assume that there is only a 50 $\mu$m space between the square narrow channel and the membrane.

Two independent straightforward numerical simulations (Finite elements with Comsol Multiphysics and Gauss-Seidel calculations implemented in Matlab) allowed us to estimate the viscous losses and always yielded similar results. 

Simulations of scenario A suggest that viscous losses amount to 0.25 W/m$^2$ for a geometry filled with only water. However, this result underestimates the dissipation because it neglects the effect of the porous capacitive electrode material. Following the work of Veerman et al.\cite{veermanReverseElectrodialysisValidated2011} and Vermaas et al. \cite{vermaas2011doubled}, we estimate the dissipation in the porous medium by considering that it is around 120 times the dissipation in a Poiseuille flow of water at the same flow rate and geometry.  This results in an estimated dissipation of 30 W.m$^{-2}$ for the given flow rate. It is very likely that the dissipation is even more important, as we did not take into account the inertial effect although the Reynolds number is estimated to be around 450. 

Simulations of scenario B yield viscous losses figures of up to 1.4 W.m$^{-2}$ for a Poiseuille flow of only water in the geometry, which corresponds to 170 W.m$^{-2}$ for the flow of water through the porous material of the electrode following the approach described above\cite{veermanReverseElectrodialysisValidated2011,vermaas2011doubled}.

If this dissipation is accurate, the authors would consume energy instead of producing 5 W.m$^{-2}$, and the viability threshold would not be met. We will revisit this point in more detail later when we will address the scalability, but it is a crucial aspect that should be addressed in the article.
Another concern is that, while the authors focus on the harvested power, they compare it directly with comprehensive studies that present a more realistic overall efficiency and account for viscous losses. As a result, their comparison may not fully support their claim of superior performance.

\section{Scalability of the experimental methodology}
\label{scalability}
The second key issue concerns the scalability of the experimental methodology, especially with regard to the fluidic switching system. The authors used a switching time that is significantly shorter than those typically reported in the literature. Although such short cycles naturally lead to high instantaneous power outputs, they are rarely (or never) used because they are entirely unrealistic for practical applications. In a real-world setting, even for a moderately sized system, such as a portable energy device (e.g., a battery-like case), rapid switching would pose significant engineering challenges. The complexity of the required control systems, electronics, and fluidic setup would result in prohibitive costs, making implementation impractical outside of highly specialized niche applications. 

If we calculate the power from the experimental results of $E$ and $I$ in Figure 2a in their paper, the power should be around 1 mW. The issue of scaling up needs to be considered, particularly since one of the usage conditions specifies that the filling time must be much shorter than the inversion period. Given that the concentration change periods are 3 seconds, this would require filling times of 0.1 seconds.
 The speed required to maintain a filling time of less than 0.1 s increases linearly with the length of the cell. Therefore, when the length of the membrane is multiplied by 10, the power dissipated per unit area is multiplied by 100, while the power produced per unit area remains unchanged. Therefore, it is not possible to increase the length of the cell to produce more energy. The only way to increase the output is to increase the width of the device and build cells that are 10 meters wide and 2 centimeters long to produce 1 watt, or to use 1000 cells to produce 1 watt. Being able to control valves that can feed such systems in such short times seems like an experimental gamble, outside of the current technological reality. 
Very large systems will most likely be complex to manage in terms of flow reversals. Therefore, it seems that the most reasonable solution lies in the upscaling of cells measuring 10 cm in width and 2 cm in length. This compartmentalization (a concept often used by nature) will require the use of 500 valves per square meter of membrane. The cost of an electric valve is significantly higher than 1 euro, which will greatly alter Post's techno-economic analysis\cite{postImplementationReverseElectrodialysis2010a}. The Capex of a device of 200 kW jumps from 98,000 euros to 20,000,000 euros, i.e. multiply it by a factor 200.

\section{Conclusions}

Taken together, the omission of hydraulic dissipation and the unrealistic switching times indicate that the proposed work may not fully address some of the key challenges currently faced by the field. As a result, the article claims regarding efficiency, performance and scalability may require further support, and the comparisons to previous work could benefit from additional refinement.

\color{black}

\section*{Author contributions}
ZM, NW, MBAF, TD, CT, AFC, AC read the article from Chapuis and Bocquet and wrote the manuscript. AC, NW, and AFC performed the Gauss-Seidel simulations, and CT and MBAF performed the finite-element simulations.

\section*{Conflicts of interest}
There are no conflicts to declare.

\section*{Data availability}

Data available on Zenodo : 10.5281/zenodo.14993066 




\balance


\bibliography{biblio} 
\bibliographystyle{unsrt} 

\end{document}